\documentclass[epjST]{svjour}

\pdfoutput = 1

\usepackage[pdftex]{graphicx,color}
\usepackage{amsfonts}
\usepackage{amssymb}
\usepackage{amsmath}

\usepackage{epigraph}

\usepackage[pdftex]{hyperref}

\let\oldmarginpar\marginpar
\renewcommand\marginpar[1]{\-\oldmarginpar[\raggedleft\tiny #1]%
{\raggedright\tiny #1}}

\newcommand{\bra}[1]{\langle#1|}
\newcommand{\ket}[1]{|#1\rangle}

\graphicspath{{figs/}}

\begin{document}


\title{Quantum annealing: the fastest route to quantum computation?}
	

\author{Christopher R.\ Laumann\inst{1,2} \and Roderich Moessner\inst{3} \and Antonello Scardicchio\inst{4,5,6,7,8} \and S.~L.~Sondhi\inst{3,4}}
\institute{Department of Physics, University of Washington, Seattle, WA 98195, USA
\and Perimeter Institute for Theoretical Physics, Waterloo, Ontario N2L 2Y5, Canada
\and Max-Planck-Institut fur Physik komplexer System, 01187 Dresden, Germany
\and Physics Department, Princeton University, Princeton, NJ 08542, USA
\and Physics Department, Columbia University, New York, NY 10027, USA
\and ITS, Graduate Center, City University of New York, New York, NY 10016, USA
\and INFN, Sezione di Trieste, Via Valerio 2, 34127, Trieste, Italy
\and on leave from Abdus Salam ICTP, Strada Costiera 11, 34151 Trieste, Italy}

\abstract{In this review we consider the performance of the quantum adiabatic algorithm for the solution of decision problems. We divide the possible failure mechanisms into two sets: small gaps due to quantum phase transitions and small gaps due to avoided crossings inside a phase. We argue that the thermodynamic order of the phase transitions is not predictive of the scaling of the gap with the system size. On the contrary, we also argue that, if the phase surrounding the problem Hamiltonian is a Many-Body Localized (MBL) phase, the gaps are going to be typically exponentially small and that this follows naturally from the existence of local integrals of motion in the MBL phase.}


\maketitle	





%

\section{Introduction: the original proposal by Farhi et al.}

\setlength{\epigraphwidth}{4in}
\epigraph{
The first branch, one you might call a side-remark, is, Can you do it with a new kind of computer---a quantum computer?
}{R.P. Feynman, Simulating physics with computers, 1982.}

In a famous aside, Feynman posited that Nature could be simulated by a device composed of discrete quantum elements \cite{Feynman:1982gn}. 
These quantum elements, not yet known as qubits, interact locally within a space-time volume asymptotically no larger than the system to be simulated.
It was apparent to Feynman that such a system would suffer from none of the difficulties that plague classical simulations of quantum physics -- neither the sign problems of quantum Monte Carlo nor the exponential overhead of wavefunction representations. 
Rather, interacting qubits form a universal quantum simulator, more powerful than any known classical device.%
\footnote{It's interesting to note that Feynman was pretty certain that a system of locally interacting spin $1/2$ objects could simulate bosonic field theories due to the known behavior of spin waves in magnetic systems. He was not at all sure that such spin degrees of freedom could simulate fermions. The modern theory of spin liquids such as the $\mathbb{Z}_2$ gauge theory/toric code has shown that fermionic degrees of freedom can likewise be locally simulated with spins.}

Fast-forward thirty years and Feynman's theoretical universal quantum simulator has evolved into a general-purpose quantum computer\footnote{Still theoretical.}, capable of efficiently solving problems outside the usual domain of Physics. 
Shor showed that a quantum computer could factor large numbers exponentially faster than any known classical algorithm \cite{Shor:1994aa}, and Grover showed that it could find a needle in a haystack provably more efficiently than any classical search party \cite{grover1996fast}. 
These results relied on the development of specific quantum algorithms%
\footnote{Shor's and Grover's.} within the unitary circuit model of quantum computation. These algorithms solve these problems and several closely related variants effectively.

It is a fact, however, that the design and analysis of quantum algorithms has not proceeded with the same pace of its classical counterpart and, apart from these notable examples we lack a ``standard algorithm" for the solution of large classes of quantum or classical problems, the role that for example DPLL plays in classical decision problems. In this context the quantum adiabatic algorithm (QAA) came as a nice surprise.

The QAA is an alternative paradigm for quantum computation, prominently proposed by Farhi et al. \cite{farhi2000quantum} in 2000, although the idea of using quantum fluctuations to speed up annealing processes for discovering the ground states of Hamiltonians more efficietly had already been proposed by \cite{Kadowaki:1998bm,Ohzeki:2010tb} in 1998. 
It is a general purpose optimization algorithm, whose goal is to determine the ground state properties of a ``problem'' Hamiltonian. 
In its original formulation, a large transverse field is applied to the system of qubits representing the problem and the system is prepared in the (simple) ground state of the field Hamiltonian. 
The algorithm proceeds by slowly turning off the external field until only the problem Hamiltonian remains. 
If this procedure is slow enough, the systems remains in its adiabatic ground state and the final state will correspond to the ground state of interest with probability close to 1.

A bit more precisely, the QAA corresponds to quantum evolution with the time-dependent Hamiltonian,
\begin{align}
	H(t) = s(t) H_p + (1-s(t)) H_b
\end{align}
where $H_p$ is the problem Hamiltonian, $H_b$ is the beginning (field) Hamiltonian and $s(t)\in[0,1]$, $s(0)=0$, $s(T)=1$ specifies an annealing protocol. 
In the simplest case, $s(t) = t/T$ specifies a linear interpolation over time $T$ in Hamiltonian space, but in many applications\footnote{Such as the solution of Grover's problem by adiabatic computation.} it is advantageous to optimize $s(t) = f(t/T)$. 
The dependence of $T$ on the system/problem size $N$ defines the efficiency of the QAA on the class of problem Hamiltonians $H_p$ at hand.

In order to guarantee that the ground state of $H_p$ is found with finite probability, $T$ must be large enough to satisfy the adiabatic theorem,
\begin{align}
	T \gg \hbar \max_{s} \frac{|V_{10}(s)|}{\Delta(s)^2}
\end{align}
where $V_{10} = \bra{1}\frac{\partial H}{\partial s}\ket{0}$ is the matrix element between the instantaneous ground and first excited states of $H(s)$ and $\Delta = E_1 - E_0$ is the corresponding energy gap.
Both quantitites depend on $s$. 
This expression makes clear the importance of the spectral gap for the success of the QAA, which is why much of the literature has focused on the scaling of the minimum gap. 
As we shall see, the matrix elements $V_{10}$ may also play an important role in determining the efficiency of the algorithm.

The QAA immediately appeals to the working Physicist. Tuning fields slowly as control `knobs' during the investigation of an experimental system is, of course, a part of the day-to-day routine in many labs. 
Morever, adiabatic passage is a well established technique for manipulating individual quantum optical systems.
The QAA bootstraps that understanding to the direct simulation of problem Hamiltonians.
It took quantum Computer Scientists only a few years to warm to the idea of robust computation by continuous evolution. 
They soon established that adiabatic evolutions could be used to simulate the evolution of quantum circuits (and vice versa) and thus that a suitably generalized QAA constitutes a completely universal model for quantum computation \cite{Aharonov:2008p7450}. 

In practice, the complexity theoretic universality of the QAA may not be nearly as important as the ``one size fits all'' scheme it provides for attacking classical optimization problems. 
In this regard, the QAA shares many features with classical simulated annealing (CSA) \cite{Kirkpatrick:1983p771}.
In CSA, the temperature of a classical Monte Carlo simulation of the problem Hamiltonian is slowly tuned to zero, ideally annealing the system to its ground state. 
Both algorithms can be readily applied to essentially arbitrary optimization problems as is. 
They both proceed by slowly turning down the fluctuational dynamics, quantum or thermal, which allow the system to explore phase space.
There are no guarantees that these fluctuations will succeed in evading metastable traps and finding the true ground state. 
Indeed, the theory of NP-completeness \cite{Arora:2009zv} essentially guarantees the contrary: 
there are classes of intractable optimization problems on which CSA and the QAA must surely fail.
Nonetheless, simulated annealing has become a workhorse of practical optimization and one hopes that the QAA will play a similar role.

After this paean to classical simulated annealing, why should we consider the QAA at all? 
As we know neither is likely to solve all NP-complete problems, the question amounts to whether there are practical problems on which we expect the QAA to perform materially better than CSA. 
This is a difficult question, as it requires understanding the physical mechanisms that underly the success or failure of each heuristic algorithm and whether those mechanisms arise in particular problem domains. 
We will return to this in more detail below.
Broadly, however, many practitioners invoke quantum tunneling as a more effective means of exploring configuration space than thermal fluctuations. 
The intuition is based on the following gedanken-optimization problem in which a single particle on a line is trapped in a local potential minimum behind a barrier of height $\Delta E$ (for small $T$ we assume the free energy $F\simeq E$) and width $w$. 
In CSA at temperature $T$, the particle escapes the trap when a sequence of like-minded stochastic fluctuations combine to provide the energy necessary to escape. 
This occurs with an exponential Arrhenius timescale 
\begin{align}
	\tau_c \sim e^{\Delta E/T}. 
\end{align}
We now turn off thermal fluctuations $T$ in favor of quantum fluctuations governed by the inverse mass $1/m$. 
The quantum particle tunnels with a rate $\sim e^{S}$ where $S$ is the action of the path tunnelling underneath the barrier. 
The WKB decay length $l \approx 1/\sqrt{2 m \Delta E}$ for a particle tunneling under a barrier of height $\Delta E$. This produces an escape timescale exponential in the width $w$,
\begin{align}
	\tau_q \sim e^{w \sqrt{m\Delta E/2}}.
\end{align}
So if $w=O(1)$, this gives $\tau_q\sim e^{\sqrt{\Delta E}}$ while $\tau_c\sim e^{\Delta E}$, which suggests that quantum tunneling is more effective at getting past high barriers $\Delta E$ than thermal fluctuations. 
Of course, quantum tunneling is less efficient if those barriers are also parametrically wide. 

In the remainder of this review, we focus on the application of the QAA to classical combinatorial optimization problems. 
As these problems typically have a high dimensional phase space with a complicated energy landscape, the intuition garnered from the particle on a line example is at best heuristic. 
Moreover, the estimates above address the escape time under fixed Hamiltonian dynamics -- not the likelihood of avoiding traps during annealing from the large fluctuation regime.
In order to make more detailed progress, practitioners have turned to case studies of the QAA applied to various model problems, with the goal of extracting generalizable success and failure mechanisms.
Unfortunately, most of the crispest results have been failure mechanisms. 
In Sec.~\ref{sec:reviewstudies}, we summarize the known failure modes from a physics perspective.
In the following two sections, we expand on two of the arguments that we find most interesting: the role of thermodynamic phase transitions and that of many-body localization (MBL) in frustrating the QAA.
Many-body localization is the interacting generalization of Anderson localization and corresponds to a dynamical phase of disordered systems in which transport is completely inhibited. 
That MBL may play a starring role in frustrating the QAA applied to random combinatorial optimization problems came as a surprise to many researchers a few years ago \cite{Altshuler:2009tf,Altshuler:2010ct}, and remains a contentious issue \cite{Farhi:2010tc,knysh2010relevance}.
Here, we will content ourselves with reframing the arguments leading from MBL to the failure of QAA in terms of local integrals of motion. 

Before continuing, we should note that perhaps the most commercially practical reason to study the QAA is that D-Wave \cite{Boixo:2014aa} claims to have built a special purpose quantum computer whose sole capability is the QAA, and Google bought one. 
The D-Wave device allows for the semi-quantum simulation of an arbitrary Ising model on the `chimera' graph subject to a tunable transverse field. 
We call the simulation semi-quantum as it is known that there is a great deal of extrinsic noise in the system on the time scales of the operation of the QAA \cite{Shin:2014wf,Dickson:2013bv,Boixo:2013ha}. 
Thus, it may be better to view the current device as a low temperature field annealer. 
In any event, the jury is out on both the `quantumness' of the machine and the asymptotic efficency of its semi-quantum annealing of hard optimization problems. 


\section{Case studies of classical difficult problems}
\label{sec:reviewstudies}

In this section, we briefly review the physical mechanisms which are known to frustrate the QAA, especially in its application to random ensembles of hard optimization problems. For a more exhausting discussion from the perspective of spin glass theory, see \cite{Bapst:2013bn}.

In principle, the quantum adiabatic algorithm can be used to search for the ground state of any Hamiltonian system. 
This includes quantum Hamiltonians which encode QMA-complete problems \cite{kitaev2002classical,Arora:2009zv} such as local hamiltonian \cite{kitaev2002classical} and quantum satisfiability \cite{Bravyi:2006p4315}. 
Relatively little is known about the typical behavior of the QAA for these quantum problems as they can not be simulated to any appreciable size.
Indeed, even extending equilibrium notions such as phase transitions and `free energy landscapes' to these models is an immature undertaking \cite{laumann2010product,laumann2010random,Hsu:2013kn}. 

Luckily, most of the optimization problems of human interest correspond to classical (diagonal) Hamiltonians. 
These have naturally attracted much more attention from researchers who have thus uncovered much more about both the equilibrium and dynamical properties of `typical' ensembles of such problems. 
Indeed, the first case study of the QAA explored its application to random ensembles of \emph{classically difficult} NP-complete problems \cite{farhi2001quantum}. 
These initial numerical studies of the performance of the QAA in solving Exact-Cover (one of the NP-complete problems) instances seemed to show polynomial scaling of the gap \cite{farhi2001quantum}.
Successive studies \cite{farhi2008make,PhysRevA.86.052334,Young:2008bk,Young:2010ct} have provided compelling evidence in favour of the exponential scaling of the gap for both some classically easy (XORSAT which is in P), and difficult problems (3-SAT and MAX-CUT, two NP-complete problem). See also the review by Hen and Young in this volume.

There are a number of physical mechanisms underlying this failure of the QAA to effectively solve these optimization problems. 
The most well understood arise in extremely non-local Hamiltonians, which can often be solved exactly. 
Quantum first order transitions which provably frustrate the QAA arise in such non-local problems whose energy functions do not provide ``basins of attraction'' suitable to local exploration in configuration space 
\cite{Farhi:2010tc,Jorg:2008fj,Farhi:2005uv,vanDam:2001jc,Jorg:2010gi}. 
This reflects the inability of local quantum fluctuations to explore non-local landscapes effectively -- either due to extensive disorder \cite{Farhi:2010tc,Jorg:2008fj} or because of flat golf-course like landscapes with exponentially small holes \cite{Farhi:2005uv,vanDam:2001jc,Jorg:2010gi}. 
In a dual sense, golf-course like driving Hamiltonians, such as using a non-local projector onto the transverse field ground state rather than the full transverse field, also lead to provably exponentially small gaps   \cite{Farhi:2005uv,vznidarivc2006exponential,Ioannou:2007aa}.
Some mean-field-like ferromagnetic models with infinite range interactions exhibit exponentially small gaps on their first order lines \cite{Bapst:2012fi}, but, as we will discuss further below, one should be careful assuming a connection between thermodynamic transitions and gap closing \cite{Laumann:2012hu,Tsuda:2013hf}.

In models with local energetics on bounded degree interaction graphs, the situation is less clear. 
Thermodynamic calculations within replica theory \cite{Smelyanskiy:2004ck,Knysh:2006p9709,Knysh:2008dk} and quantum cavity theory \cite{laumann:134424} suggest random quantum first order transitions persist in at least some local models \cite{Jorg:2010ys}, although QMC data is inconclusive \cite{Young:2008bk}. 
Controversial work suggests that Anderson localization may arise in configuration space when quantum fluctuations are very weak, leading to `perturbative crossings' and exponentially small gaps near the endpoint of the evolution \cite{Knysh:2010un,Farhi:2009vh,Altshuler:2010ct,Amin:2009kf}.
Heuristic arguments assuming the presence of `clustering' of pure states in a glassy phase suggest that such crossings may arise throughout an extended regime of the adiabatic evolution \cite{Foini:2010gj}; such crossings may even have been observed in QMC \cite{Young:2010ct}. 
We will return to localization in more detail below and present a new heuristic derivation of its dire consequences for the QAA.

In disordered, geometrically local optimization problems in 1D, Griffiths-like effects may arise in which large local regions order before the whole \cite{BenReichardt:2004vd,PhysRevB.51.6411}. It is unlikely that such Griffiths effects play an important role in higher dimensions or on long-range interaction graphs, as they tend to be most important in very low dimensions.

\section{Small gaps from critical points}
%
%


Critical slowing down is the name given to the phenomenon that at
second order phase transitions, not only do (spatial) correlation
lengths but also (temporal) correlation times diverge \cite{chaikin1995}. This means in
particular that, in order for a simulated annealing algorithm to
remain in equilibrium through a second order phase transition, the
rate of cooling must vanish as the critical point is
approached. 

From a complexity point of view, this is not as bad as it sounds. 
Since at second order phase transitions quantities vary as power laws with the distance from the critical point, the sweep rate will only need to be power-law small in the system size. 
This is what is known as `easy' in complexity theory. 

At second order quantum phase transitions, the corresponding
expectation is for gaps in the many-body spectrum to be algebraically
small. 
Indeed, the idea of a dynamical critical exponent states that a
gap, $\Delta$, should vanish according as a power law in the
separation, $g-g_c$ from the critical point $g_c$:
\begin{equation}
\Delta \sim \xi^{-z} \sim (g-g_c)^{\nu z},
\end{equation}
where $z$ is known as the dynamical critical exponent relating the correlation
length $\xi$ to the gap $\Delta$. As an aside, notice that for quantum
systems, which are specified by a Hamiltonian as well as by the
commutation relations between the operators it contains, the dynamics
is fully specified, whereas for classical systems, a given Hamiltonian
may be supplemented with a range of different dynamical prescriptions. 
This may in particular lead to different dynamical scaling relations. 

The power law form is then bequeathed to the finite-size scaling
properties. Thus, as far as the gap $\Delta$ is concerned, we are
still in an `easy' regime.

A priori, things look considerably more grim for the performance of
the QAA in the case of first-order phase transitions. 
Here, the expectation is that gaps will be exponentially rather than
algebraically small in system size. The basic intuition behind this
can be formulated in a number of ways. The first stems from the fact
that correlation lengths remain finite at such transitions. Local
operators should therefore have expectation values which are
exponentially small in the number of correlation volumes contained in
the system, and level crossings should be correspondingly
exponentially small in size. Another goes along with the observation
that at a second order transition, gapless excitations in the form of
a soft mode exist, e.g.\ when a spin wave or triplon dispersion become
gapless in a magnet. These gapless excitations underpin the algebraic
finite-size gap if they have a power-law dispersion, as is
generically the case when a dispersion minimum goes through zero
at the soft mode transition. 
At first-order transitions, no such modes
exist, and hence there is no mechanism for generating polynomial gaps. 

This expectation is indeed regularly fulfilled, and in the course of
time has acquired the status of a folk theorem. However, it turns out
to be incorrect. The reason is basically the following. Whereas the
above arguments on soft modes involved excitations in a hydrodynamic
sense -- spin waves in a Heisenberg magnet being related to
spin-rotational symmetry breaking -- it is in fact also possible to
have `accidental' excitations which are due to, e.g., `inappropriate'
choices of boundary conditions. 

In this vein, let us consider \cite{Laumann:2012hu} the case of an
antiferromagnetic Ising spin chain in a transverse field $\Gamma$. 
Let us also add a staggered field, $h$, to
this, which prefers one particular of the two possible Neel states 
as ground states for small $\Gamma$. 
\begin{equation}
H_{afm}=\sum_i \sigma_i^z\sigma_{i+1}^z + (-1)^i h \sigma_i^z + 
\Gamma \sigma_i^x
\end{equation}

As the staggered field changes sign, all spins need to be flipped so
that the spin state is reversed from one ground state to the
other. This is in fact a thermodynamic first-order transition,
trivially related to that of a ferromagnetic Ising model in a
longitudinal field at zero temperature.

The way the system can dynamically flip all the spins is by sweeping a
domain wall across the system. For a chain of even length, this
process requires the generation of a domain wall, which then needs to
hop $L$ steps until all spins are flipped. Due to the excitation
energy of the domain wall, the resulting action, and hence the
splitting between the ground states, is exponentially small in $L$ --
as expected for a first-order transition.

However, what happens for a spin chain of odd length? In that case,
there always has to be a domain wall somewhere in the system, and the
staggered field will chose a preferred location for it.  However, as
the sign of the field is now changed, it no longer is necessary to
generate a domain wall before sweeping it across the system -- it's
already there in the ground state. What one finds instead is a hopping
problem for the domain wall, which at the transition point has a
cosine-shaped dispersion with a minimum at zero energy. Hence, the
finite-size gap is again algebraically small, and we have an avoided
crossing which is much bigger than exponential. So, for some
first-order transitions, the QAA does not fail!

The lesson from this is that the many-body finite-size gap is not the
same as that of the local hydrodynamic modes usually considered in critical phenomena. 
In the simplest examples, this happens in the same way that
choosing antiperiodic boundary conditions for an Ising ferromagnet
will not change the dispersion of its excitations, but can give rise to gapless excitations of the defect which is unavoidably present.
More generally, the many-body finite size gap at first order transitions may exhibit extremely complicated size dependence due to commensuration effects. 
For example, the first order transition in the long-range ferromagnetic XY model exhibits a finite-size gap scaling polynomially, exponentially or even factorially small depending on the precise sequence of system sizes taken in the thermodynamic limit \cite{Tsuda:2013hf}.

%

\section{Small gaps from the integrals of motion of MBL}

In recent years our understanding of quantum systems with quenched disorder has greatly improved thanks to research on many-body localization (MBL). 
The concept of localization originated in the work of Anderson on disordered  lattices \cite{Anderson:1958p7531}; the question of its stability in the presence of  interactions has been a theoretical pastime ever since \cite{Fleishman:1980cz}. 
Modern computational power, the development of quantum optical systems where these questions might be probed experimentally and the perturbative analysis of Basko, Aleiner and Altshuler \cite{basko2006metal} have motivated a recent explosion of interest and progress in understanding the interacting (hence `many-body') form of localization. 

Now it appears that MBL phases appear generically in the strong disorder region of quantum spin systems, ranging from one-dimensional  \cite{oganesyan2009energy,pal2010many,de2013ergodicity} to fully connected models \cite{buccheri2011structure,laumann2014many}. 
The MBL phase can be non-perturbatively characterized by the existence of an extensive collection of many-body local integrals of motion \cite{huse2013phenomenology,serbyn2013local,imbrie2014many,ros2014integrals}.
The presence of these conserved quantities inhibits transport on large distances in the system, protect $\mathbb{Z}_2$ order against perturbations at infinite temperature \cite{huse2013localization} and slow down considerably the spread of entanglement \cite{vznidarivc2008many}.

All these features suggest that the adiabatic algorithm will indeed run into trouble in MBL phases. 
Morally, as quasi-static quantum fluctuations are unable to change the local integrals of motion, they will be unable to explore phase space effectively. 
In the context of random combinatorial optimization problems, the MBL regime would be a phase extending around the problem Hamiltonian of interest, causing difficulty for the QAA as it comes into the homestretch. 

This scenario was first explored by Altshuler, Krovi and Roland (AKR) in \cite{Altshuler:2010ct} where it was argued that MBL allows for the perturbative treatment of level crossings near the classical endpoint of the QAA and that these in turn lead to exponentially small gaps. 
Indeed, in the most recent numerical studies like \cite{PhysRevA.86.052334}, the fastest decreasing gap found by quantum Monte Carlo techniques appears well within the spin-glass region, rather than simply at the transition into it.
We note that we generically expect the spin-glass phase, as predicted by statistical mechanics, to lie within the dynamical MBL region, as this is the case in the only model in which both phases have been analyzed \cite{laumann2014many}.

Based on a case study of random instances of the problem EXACT COVER 3 (EC3), AKR conjecture that random NP-complete optimization problems generically exhibit MBL phases and small crossings in the vicinity of the problem Hamiltonian. 
In the following, we will review their arguments and reframe them in terms of integrals of motion. 
This will lead us to the observation that the mechanism for avoided level crossings identified by AKR can be extended throughout any MBL phase, rather than simply near the endpoint. 
However, there are many conjectural assumptions that go into this argument (just as with AKR's original work) which may or may not be satisfied depending on the nature of the decision problem at hand. 
We will return at the end to which aspects we feel are on solid ground and which need much more work.

Let us now review these arguments in the context of EXACT COVER 3. The cost function of the problem EC3 is a positive integer
\begin{equation}
E_C(x)=\sum_{a=1}^M(x_{i_a}+x_{j_a}+x_{k_a}-1)^2,
\end{equation}
where $M$ is the number of clauses and $x_i=0,1$ are the bit assignments and if $E_C(x)=0$ we have found a satisfying assignment. As a function of $\alpha=M/N$ there is a transition at $\alpha_c\simeq0.62$, where the probability to have a satisfying assignment of the random formula goes from 1 for $\alpha<\alpha_c$ to 0 for $\alpha>\alpha_c$. By going from $x_i$ to $\sigma^z_i=x_i-1/2$ and introducing the conjugate operator $\sigma^x$ we can write the adiabatic Hamiltonian
\begin{equation}
H(t)=s(t)E_c(\sigma)+(1-s(t))\sum_{i=1}^N\sigma_i^x,
\end{equation}
where
\begin{equation}
E_c(\sigma)=-\sum_i h_i \sigma_i-\sum_{i,j}J_{ij}\sigma_i\sigma_j,
\end{equation}
is a classical Hamiltonian with integer coefficients.

For $1>s>0$ one can introduce the parameter $\lambda=(1-s)/s$ and consider perturbation theory in this parameter. When perturbation theory converges one can link in a one-to-one way an eigenstate $\ket{n}$ to a spin configuration $\ket{\sigma}$ and write down energies of eigenstates in a series
\begin{equation}
\label{eq:EnL}
E_n(\lambda)=E_C(\sigma)+\sum_{m=1}^\infty \lambda^{2m}F^{(m)}(\sigma).
\end{equation}
This expression can be used to find an avoided crossing of two levels for which $E_C(\sigma_1)\neq E_C(\sigma_2)$ but for a certain $\lambda$, $E_2(\lambda)\simeq E_1(\lambda)$.

AKR claim that this happens for $\lambda\gtrsim \lambda_*\sim 1/N^{1/8}$ and $\lambda\lesssim \lambda_{cr}\sim1/\ln N$. This latter value of the critical ``hopping" $\lambda_{cr}$ follows from a parallel with localization on the Bethe lattice \cite{abou1973selfconsistent}. 
Taking the difference between $E_2$ and $E_1$ of (\ref{eq:EnL}) and considering the behaviour of the typical values of $F_{1,2}$ obtained from the numerics, they find that the difference in energy between two eigenstates pertaining to different spin configurations (at $\lambda=0$) is
\begin{equation}
|E_1(\lambda)-E_2(\lambda)|=\sqrt{N}\sum_{m=2}^\infty \lambda^{2m} f^{(m)},
\end{equation}
with $f^{(m)}=(F^{(m)}_1-F^{(m)}_2)/\sqrt{N}$ a random variable of $O(1)$. 
The sum starts from $m=2$ rather than $m=0$ because the lower order terms are independent of the configuration and cancel out.
By starting from two satisfiable configurations and adding a clause which leaves only one of the two satisfiable, we get the equation $|E_1(\lambda)-E_2(\lambda)|=O(1)$ for the location of the level crossing, which gives $\lambda_*\sim N^{-1/8}$ quoted before.

This avoided crossing gives an exponentially small gap if the Hamming distance between $\sigma_1$ and $\sigma_2$ is $d(\sigma_1,\sigma_2)=\nu(\alpha)N=O(N)$. 
That random EC3 possesses such macroscopically distinct global ground states is known from classical statistical mechanical studies \cite{biroli2000variational}. 
From this avoided crossing, one obtains a gap
\begin{equation}
\Delta\sim e^{-\frac{\nu}{8} N\ln(N/N_0)}
\end{equation}
where $N_0=O(1)$. This is even \emph{smaller} than exponential.

A crucial point of the analysis of \cite{Altshuler:2010ct} is the convergence of perturbation theory for the energy $E_n(\lambda)$, which is claimed on the basis of the existence of an MBL region for $\lambda<\lambda_{cr}$. 
The existence of a localized region is probably on safe ground, although the value of $\lambda_{cr}$ is far from certain, since the parallel with the Anderson model on the Bethe lattice requires several crucial assumptions. 
One of these, statistical independence of the energies of neighbouring configurations, is certainly violated. 
Another assumption, the absence of loops in the hypercube of configurations, is also certainly violated.

The reasons to believe that a $\lambda_{cr}$ exists that separates the MBL from the ergodic phase should be based on considerations more akin to those presented in \cite{basko2006metal,ros2014integrals,laumann2014many}, where the assumption of the independence of the energies of neighbouring configurations and/or the absence of loops is not made.

Even leaving the question of the existence of an MBL region however, a criticism was posed to \cite{Altshuler:2010ct} in \cite{knysh2010relevance}. The authors of \cite{knysh2010relevance} claim that the level splitting is not exponential and blame the fact that the system size analyzed in \cite{Altshuler:2010ct} are too small. In their analysis the avoided crossing occurs at $\lambda^*\sim 1\gg \lambda_{cr}$ for sufficiently large $N$, therefore invalidating the use of perturbation theory.

However we find that this objection may not be as fatal as it looks at first sight, and in the following we propose a possible line of argument for tightening the connection between MBL and computational hardness which suggests that an appropriate perturbation theory may be constructed `locally' in $\lambda$. We hope that this will stimulate further detailed study of this conceptually important aspect of both MBL and the QAA.

If there is an MBL phase for $\lambda<\lambda_{cr}$  then perturbation theory in the hopping $\lambda$ should converge (with large probability). Then, the question of the smallness of the gap is discussed most naturally in the set-up of the integrals of motion for MBL.
Let us now recall how these are defined \cite{huse2013phenomenology}. Starting from the MBL phase of a generic spin Hamiltonian, which we restrict to the form relevant to the QAA:
\begin{equation}
H=H_p(\sigma_z)+\lambda H_b(\sigma_x).
\end{equation}
In the whole MBL phase, for all $\lambda<\lambda_c$, one can perform a convergent series of local unitary rotations \cite{huse2013phenomenology,imbrie2014many} which sums up to a unitary $U$ which at the same time brings
\begin{equation}
U^\dag\sigma^z_i U=\tau^z_i,
\end{equation}
and
\begin{equation}
\label{eq:lbitham}
H=H'(\tau_z)=-\sum_{i}h_i(\lambda)\tau^z_i-\sum_{i,j}J_{ij}(\lambda)\tau^z_i\tau^z_j-\sum_{i,j,k}J_{ijk}(\lambda)\tau^z_i\tau^z_j\tau^z_k+...
\end{equation}
where the couplings $J$'s have short range (exponential decay with MBL length $\xi$). The transformation $U$ can be found as a power series in $\lambda$ (for a model of spin chain with disorder this transformation has been proved to exist in \cite{imbrie2014many})
\begin{equation}
U=\mathbb{I}+\lambda u^{(1)}+\lambda^2 u^{(2)}+...\ ,
\end{equation}
which is convergent for any $\lambda<\lambda_c$.
The $l$-bits $\tau$ ($l$ is for ``localized") can be used to label the exact $2^N$ MB eigenstates with a bit string
\begin{equation}
\ket{\Psi_n}=\ket{\tau_1,...,\tau_N}\equiv\ket{\vec{\tau}}.
\end{equation}
The couplings $h,J$'s are all functions of $\lambda$, which reduce to the classical hamiltonian for $\lambda=0$. Usually for the classical hamiltonian one has integer $h,J$ and only up to a fixed number of spins interaction $J_{i_1,...,i_n}= 0$ for $n>K$ in K-SAT for example (counterintuitively, $K=2$ for EC3 as the 3-body interaction can be expressed as a quadratic form).

Since the unitary rotation $U$ can be obtained as a convergent series of local unitary operators \cite{imbrie2014many} for $\lambda<\lambda_c$, the resulting hamiltonian $H'(\tau_z)$ has couplings which depend analytically on $\lambda$ (with radius of convergence at least $\lambda_c$) and decrease exponentially with the distance between the indices $i,j$ etc.

For example, considering the case in which only $h_i$ and $J_{ij}$ are $\neq 0$ for $\lambda=0$ the couplings have the form
\begin{eqnarray}
h_i(\lambda)&=&h_i^{(0)}+\lambda h_i^{(1)}+\lambda^2 h_i^{(2)}+\lambda^3 h_i^{(3)}+...\nonumber\\
J_{ij}(\lambda)&=&J_{ij}^{(0)}+\lambda J_{ij}^{(1)}+\lambda^2 J_{ij}^{(2)}+\lambda^3 J_{ij}^{(3)}+...\nonumber\\
&\vdots&\nonumber\\
J_{i_1,...,i_n}(\lambda)&=&\lambda^b(J_{i_1,...,i_n}^{(1)}+\lambda J_{i_1,...,i_n}^{(2)}+...),
\end{eqnarray}
where $b>0$ and grows with $n$.

The question of the smallness of the gap is now turned into the question of what are the gaps of the \emph{classical} hamiltonian $H'$. We will see that it is easy for this Hamiltonian to have exponentially small gaps. In particular, small gaps are natural if, in the adiabatic transformation, the value of an \emph{extensive} number of local integrals of motion change from $\lambda=\lambda_c$ to $\lambda=0$. In the following we try to explain this mechanism in detail.
 
\begin{figure}[htbp]
\begin{center}
\includegraphics[width=0.47\columnwidth]{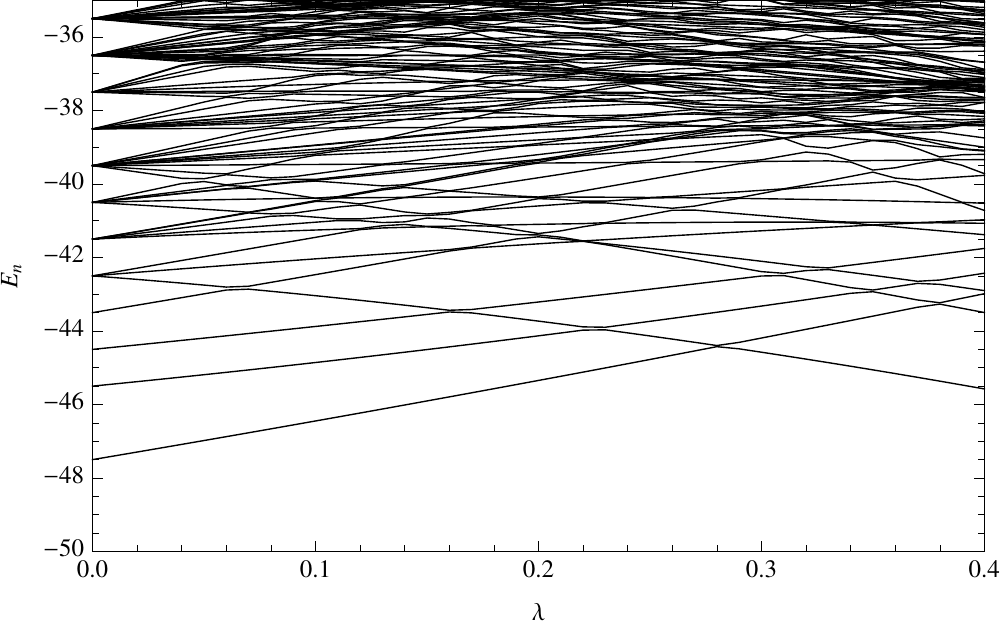}
\caption{Exact crossing for a $N=15$ spin model (\ref{eq:lbitham}) with random integer fields and 2-spin couplings, which evolve smoothly with $\lambda$. Only the lowest 100 levels are shown. The level crossing the ground state at the smallest $\lambda\simeq 0.28$ has Hamming distance $5=L/3$ from the ground state. The $l$-bits assignments of the ground state change extensively from $\lambda=\lambda_c$ to $\lambda=0$.}
\label{default}
\end{center}
\end{figure}

First of all one should notice that the integrals of motion $\tau_a$ are \emph{not conserved} by the adiabatic evolution. This is because the commutation rules
\begin{equation}
[H(t),\tau_a(t)]=0
\end{equation}
of $\tau$ with a time-dependent Hamiltonian \emph{do not imply} that
\begin{equation}
[\text{T}e^{-i\int_0^t dt' H(t')},\tau_a(0)]=0,
\end{equation}
where T is the time-ordered product.

As one reduces $\lambda$ from large to small in the QAA, let us assume that one enters an MBL phase at $\lambda=\lambda_c$. We make the further hypothesis that \emph{the entire spectrum} is MBL and that the Hamiltonian (\ref{eq:lbitham}) describes all the eigenvalues and eigenstates.
Assuming the transition itself has polynomially small gaps (as seen for example in \cite{PhysRevA.86.052334}), just after the transition the state is with high accuracy in the ground state of $H(\lambda_c)$. This corresponds to some assignments of the $\tau$'s
\begin{equation}
\Psi(\lambda_c)=\ket{\tau_1(\lambda_c),...,\tau_N(\lambda_c)}=\ket{\vec{\tau}(\lambda_c)}.
\end{equation}
However, since the integrals of motion are not conserved in the arbitrarily slow adiabatic evolution, the classical ground state can have assignments of the integrals of motion which are \emph{extensively} different from those of the quantum ground state at $\lambda_c$:
\begin{equation}
\label{eq:extensivedist}
d(\vec{\tau}(0),\vec{\tau}(\lambda_c))=O(N).
\end{equation}
The last bit which we are left to prove is that the crossing between states which are $O(N)$ spin-flips apart is exponentially small in $N$. 

For this we need the fact that the unitary transformation that sends $\sigma^z\to\tau^z$ also sends $\sigma^x\to\tau^x$, where the commutation rules are preserved
\begin{eqnarray}
&U^\dag\sigma_i^{x,y}U =&\tau^{x,y}_i, \\
&[\tau^a_i,\tau^b_j]=& i\delta_{ij}\epsilon_{abc}\tau^{c}_j.
\end{eqnarray}
Therefore one consider the perturbation of $\lambda\to\lambda+\delta\lambda$, with $\delta\lambda=O(\lambda)$ and $\lambda<\lambda_c\ll 1$. Then we can write
\begin{equation}
H(\lambda+\delta\lambda)=H(\lambda)+\delta\lambda\sum_{i}\sigma^x_i,
\end{equation}
which can be written, going from $\sigma$ to $\tau$'s:
\begin{eqnarray}
H(\lambda+\delta\lambda)&=&H(\lambda)+\delta\lambda\sum_{i}\tau^x_i+O(\lambda\delta\lambda)\nonumber\\
& = & -\sum_{i}h_i(\lambda)\tau^z_i-\sum_{i,j}J_{ij}(\lambda)\tau^z_i\tau^z_j-\sum_{i,j,k}J_{ijk}(\lambda)\tau^z_i\tau^z_j\tau^z_k+...+ \nonumber\\
&+&\delta\lambda\sum_{i}\tau^x_i+O(\lambda^2)
\end{eqnarray}
where we should keep only the terms in the hamiltonian which are of $O(\lambda)$ or larger. If we want, we could to re-absorb the $\delta\lambda$ term we can, by redefining the coefficients $h_i(\lambda), J_{ij}(\lambda),... \to h_i(\lambda+\delta\lambda), J_{ij}(\lambda+\delta\lambda),...$ etc.\ and $\tau^a(\lambda)\to\tau^a(\lambda+\delta\lambda)$.

By changing $\delta\lambda$ from 0 to $\delta\lambda$, assuming there is an avoided level crossing somewhere in the range $[0,\delta \lambda]$ keeping the Hamiltonian in this form, with the assumption (\ref{eq:extensivedist}), it follows easily that the avoided crossings generate exponentially small gaps. In fact, the perturbation $\delta\lambda$ flips only one $\tau_i$ at a time, and if the crossing has to be avoided, one needs to apply the perturbation $d=d(\tau(\lambda_c),\tau(0))$ times, generating therefore a gap proportional to the matrix element:
\begin{equation}
\Delta\simeq \delta\lambda^{d}\prod_{n=1}^d\frac{1}{E_0-E_n}\propto \delta\lambda^{\nu N}\sim\lambda^{\nu N},
\end{equation}
where $\nu=O(1)$ and the product is over the order of spin flips which maximises the amplitude (some work is necessary to show that this actually gives at most an exponential contribution but this has been discussed at length in \cite{basko2006metal,ros2014integrals}).

To summarise, the existence of integrals of motion in the entire MBL region means that a slight increase of the external magnetic field by $\delta\lambda$ has the exactly same effect of applying a perturbatively small transverse magnetic field to a classical (containing only $\sigma_z$'s) spin hamiltonian: if, because of the field, a crossing occurs between states which have extensively different values of the spins/integrals of motion, then the avoided crossing is necessarily going to be exponentially small.

The question is now reduced to finding these two almost degenerate states which have extensive difference in the values of the $l$-bits integrals of motion. In the hypothesis of AKR, where the crossing is perturbative, this is a purely classical question and depends on the  classical structure of the decision problem. This structure has been studied extensively for some classical problems like XORSAT, EC, SAT etc.\ and some details of the interaction graph seem to matter more than the complexity class of the problem itself. If the crossing occurs at non-perturbative $\lambda$ (although $\lambda<\lambda_c$) as suggested in \cite{knysh2010relevance}, then the structure of the classical problem is less relevant, and one should look at the structure of the classical $H(\lambda)$ in terms of $\tau^z$'s. It is possible that the structure of $H(\lambda)$ for $\lambda$ not perturbatively small is radically different from that of $H(0)$ but only an extensive investigation can resolve this issue for the given decision problem.

We point out that this phenomenon is reminiscent of ``temperature chaos" in some spin glasses, in which at an arbitrary small change of the external parameters (usually temperature) the state of the system changes abruptly. At the change of the external magnetic field, states which are macroscopically different swap positions in the spectrum, and this originates a wealth of exponentially small gaps in avoided crossings. We note that a version of the arguments above have been applied to local perturbations in the MBL phase in parallel work \cite{Khemani:2014aa}.

Finally, on the grounds that QMA-hard (quantum) problems should be at least as refractory NP-hard (classical) problems, we conjecture further that random QMA-complete problems, such as quantum satisfiability, would likewise reside within MBL phases and exhibit ``field chaos''. 
There are limited numerical studies to support this claim \cite{Joonas:2008aa}; we believe this to be an interesting direction for future research.

\section{Conclusion}

We have described the current understanding of the behaviour of the QAA for the solution of classical decision problems like EC, XORSAT or SAT. We have discussed how the common lore linking exponentially small gaps to first order transitions and polynomially small gaps to second order phase transitions can be deceptive. We have also pointed out that a common source of exponentially small gaps is (as pointed out in \cite{Altshuler:2010ct}) an MBL region surrounding the classical point $\lambda=0$. With the help of the local integrals of motion of MBL, and an assumption on the structure of the classical spectrum, we have proved that avoided crossings in the MBL region give exponentially small gaps.

Linking MBL to the failure of the QAA and possibly using this information to improve the performance is a promising new line of research on which progress we hope we will be able to report in the near future.

\section{Acknowledgements}
A.S.\ thanks the Princeton Center for Theoretical Physics for hospitality, A.S.\ (in part) and S.L.S.\ are supported by the NSF grant PHY-1005429. A.S.\ thanks B.Altshuler, D.Huse and V.Oganesyan for continuing discussions on the nature and implications of MBL.

\bibliography{ReviewQAA}
\bibliographystyle{plain}

\end{document}